\documentstyle[12pt]{article}
\begin{document}

{\hfill  \bf Preprint Alberta-Thy-11-95.}

\begin{center}
{ \Large \bf
The Spectrum and Confinement for the Bethe-Salpeter Equation.
}\\[10mm]
\noindent
  A.Yu. Umnikov and F.C. Khanna\\[4mm]
{\em
 Theoretical Physics Institute, Physics Department,
 University of
 Alberta, \\Edmonton,
  Alberta T6G 2J1,
 \\ and
 TRIUMF, 4004 Wesbrook Mall, Vancouver, B.C. V6T 2A3, Canada}\\[4mm]

 \end{center}

\newpage

\begin{abstract}
The problem of calculating of the mass spectrum of the two-body
Bethe-Salpeter equation is studied with no reduction to the
three-dimensional  ("qua\-si\-po\-ten\-tial") equation.
The method to find
the ground state and excited states for a channel with
any quantum numbers is presented.
The problem of the confining interaction  for the Bethe-Salpeter
equation is discussed from the point of view of
formal properties of the
bound state spectrum, but with only an inspiration from the QCD.
We study the
kernel that is non-vanishing at
large Euclidian intervals, $R_E\to \infty$,
which is constructed as a special limiting case of a
sum of the covariant one-boson-exchange kernels.
In the coordinate space
this kernel is just a positive constant and
corresponds to the kernel $\propto\delta(k_E)$ in the momentum space.
When the usual attractive interaction is added,
it is found that this kernel is similar in its effect
to the non-relativistic potential in coordinate space, $V(r)$,
with $V(r\to \infty) \to V_{\infty}$.
The positive real constant $V_{\infty}$ gives
the scale that define the limit of  bound state spectrum compared to
the sum of the constituent masses, $M < 2m + V_{\infty}$.
At the same time the self-energy corrections
remove the singularities from the propagators of the constituents,
i.e.  constituents do not propagate as free particles.
Combination of these features of the solutions allows
an interpretation  of this type of interaction as
a confining one.
The illustrative analytical and
numerical calculation are presented for a model of  massive
scalar particles with scalar interaction, i.e. the "massive Wick
model".
\end{abstract}

\newpage

\section {Introduction}
The  Bethe-Salpeter (BS) equation~\cite{classic,nakan}
provides a consistent covariant description
for bound state of two fermions (or fermion-antifermion).
 In particular, a phenomenological
success has been achieved in the case of the deuteron
describing ground state properties
and various reactions~\cite{tjond,ukk}. It has been established
that BS formalism reproduces the bulk of the non-relativistic
approaches at the nucleon momenta small comparing to the
nucleon mass, $|{\bf p}| < m$, but displays a relativistic effects
at larger momenta $|{\bf p}| \geq m$ and in some of the
spin-dependent
observables. However, the relativistic effects are not too large,
since the
deuteron is essentially a non-relativistic system.

Another topical subject studied within the BS formalism  is
the bound state of the quark-antiquark systems, mesons ($q\bar q$).
Compared
to the deuteron case the $q\bar q$-bound states are
governed by a more complex dynamics and, correspondly, physics here
is richer.
Usual way to proceed here is to reduce the BS equation
to an approximate equation of the quasi-potential form
and then to use approaches and methods similar to those
which are used in the non-relativistic Schr\"odinger equation.

A number of the specific questions of principle and technical
problems must be resolved in order
to reach the same state-of-art as it
is in the case of the deuteron:
\begin{enumerate}
\item First, we need a rigorous method to obtain the
 bound state spectrum
 and the corresponding amplitudes for the BS equation.
 Today we do not find in the literature a discussion of this
 problem without making approximations.
\item Second, the fundamental problem here is a construction of
 an effective theory, which would provide the BS equation
  with  confinement. Moreover, it is not clear how the confinement
 should  manifest itself in this formalism.
\item Third problem, we would like to mention here, is the "problem
 of light quarks ($u$ and $d$)".
 For the formalism ofthe  BS equation we understand this problem
 as a description of the dynamical generation of the
 masses and creating the bound states with a mass, which is
 much larger then sum of the "bare" masses of the constituents.
 Apparently, physics of light quarks is essentially different
 of physics of heavy quarks.
 We will not discuss the special
 problems of light quarks physics in this paper.
\end{enumerate}

All these problems were discussed starting the pioneering work
of Llewellyn-Smith~\cite{LS}.
There are, of course,  methods to study the spectrum based on
the reduction of the two-dimensional integral equations to an
{\em infinite} system of one dimensional equations by means of the
hyperspherical harmonics and, then cutting off the system
at some lowest
functions
as an approximation~\cite{LS,kercoor,guth}.
 This method is valid for the tightly bound
states, with mass vanishing compared to the sum of the constituent
masses,
in which case the BS equation has an exact ${\cal O}(4)$-symmetry.
This was sufficient for early studies, where
 the mechanism of the confinement
was modeled as the binding of the heavy quarks to the
relatively light bound states.
However, it has become  clear that even for heavy quarks
the mass spectrum of the mesons lies near and even above the
sum of the constituent masses. Therefore other methods must be
employed to at the spectrum with such characteristics.

Alternative approaches to the study of the spectrum of the
relativistic
bound state is based on the reduction of the BS equation to
one of the possible quasi-potential equations and then utilizing
the Schr\"odinger-like potentials, $ V \sim \alpha r$, in the
coordinate
space~\cite{tjonqq} or its generalization in the momentum
space~\cite{grossmilana,khanna,vary}.
In spite of the obvious advantages of these
approaches, such as clear physical interpretation of interactions and
solutions with relationship to the Schr\"odinger equation
approach~\cite{jpr},
 still there is interest to study the full
BS equation, which gives direct connection to the covariant field
theory.

As to the underlying effective field theory for the relativistic
equations
of the bound states, now we can only imagine about the structure
of such a theory and its connection to the fundamental theory, QCD.
The approach based on the integral equations of the field
theory~\cite{qpr},
the set of Schwinger-Dyson
equations for propagators, the Ward-Takahashi  relations
(Slavnov-Taylor relations)
and
the Bethe-Salpeter equation
 is very promissing. The model investigations presented here
allow for a study of the properties of the analytical structure of
the quark propagator~\cite{qpr,williams},
the mechanisms of the   chiral symmetry breaking and
the dynamical generation
of the masses~\cite{grossmilana} and
the connections with the bound state
problem~\cite{munczek}
and
observables of reactions~\cite{tw}.

In the present paper we (i) consider the problem of the
spectrum of the two-body
BS equation with no reduction to the
quasipotential equation or other approximations and (ii)
attempt to construct a model for the confinement, based
on the formal properties of the kernels, but with no
direct connection
with QCD.
We present, in Section~2, a method to find the ground states and
excited states
in the channel with any quantum numbers. The
illustrative numerical calculations, aimed to
show capabilities  of the method  are presented in Section~2.3.
We consider in Section~3 the formal properties of the
kernel in the form of a sum of the ladder kernels.
 We derive
conditions when this kernel displays properties,
similar to the non-relativistic constant potential in the coordinate
space.
 The implementation
of this interaction to the single particle propagator
leads to the disappearance of the singularities in the propagator.
Using the developed method to study the bound state spectrum
we analyze the BS equation with this kernel and self-energy
corrections in Section~3.3.
Combination of properties of the solutions in this case
allows for an interpretation of the
interaction of this type as the confining interaction.

{\em Nota bene}. In this paper we often use
the name "quark" for the constituents
and the "meson" for the bound system. This is just a matter of
convenience and our concrete calculations are not based
on the QCD and do not pretend to a phenomenological
application. Moreover, the illustrative analytical and
numerical calculations are presented for a model with massive
scalar particles with scalar interactions, the "massive Wick model".
However, the methods can be  applied directly  to the BS
equation for two spinor fields interacting via a phenomenological
$q\bar q$-interaction. The spectrum of the spinor-spinor
BS equation has been studied in ref.~\cite{llwi95} with
no confining interaction, and the application of the model
with the confinement
to the meson
spectrum will be done elsewhere.

\section {The bound state spectrum of the Bethe-Salpeter equation}

\subsection{The scalar Bethe-Salpeter equation}

The BS equation for bound state of two equal mass scalar fields,
"quarks",
interacting via scalar exchanges has
a form:
{\begin{eqnarray}
\psi (p,P) =  D(p_1,p_2)^{-1}\int
\frac{d^4 p'}{(2\pi)^4}{\cal K}(p,p',P)\psi(p',P),
\label{sbse}
\end{eqnarray}
}
where $\psi(p,P)$ is the BS amplitude for the bound state with
momentum
$P$, ${\cal K}(p,p',P)$ is a kernel of
BS equation and  $D(p_1,p_2)^{-1}$ is the two-body propagator.
Generally speaking, the kernel ${\cal K}$ is a sum of the all
renormalized irreducible graphs presenting the Green's function
with two incoming and two outcoming constituent fields and
the propagator
$D^{-1}$ is product of the renormalized full single-particle
propagators.
In practice, however, both of them are usually taken in the lowest
order in the coupling constant. It corresponds to
 the ladder approximation for
the kernel with free particle
propagators.
In the simplest case
of an exchange field of mass $\mu$ and coupling constant $g$,
the ladder approximation for the
kernel  has an explicit form:
{\begin{eqnarray}
{\cal K}_{ladder}(\mu,p,p',P) = {\cal K}_{ladder}(\mu,p,p') \equiv
\frac{{\it i}g^2}{(p-p')^2-\mu^2},
\label{ladk}
\end{eqnarray}
}
The two-body propagator then reads
{\begin{eqnarray}
D(p_1,p_2)\equiv d(p_1)\cdot d(p_2) =
\left ({p_1^2-m^2}\right )\cdot
\left ({p_2^2-m^2}\right ),
\label{twoprop}
\end{eqnarray}
}
where $d(p)^{-1}$ is free one-particle
propagator, $m$ is the mass of the constituents and
$p_{1,2} = \frac{1}{2}P \pm p$ are the constituent
momenta.

Phenomenological applications of the equation (\ref{sbse})
(or more interesting spinor-spinor equation)
requires a more general form of the kernel than eq. (\ref{ladk}).
More degrees of freedom are provided by introducing
a sum of the exchanges with different parameters and different
Lorents structure, e.g. vector or pseudoscalar exchanges.
In the presence of the several (effective) exchanges with
different quantum numbers and in  higher orders kernel may
contains both attractive and repulsive terms.
Here we consider
the kernel of the form:
{\begin{eqnarray}
{\cal K}_G(p,p') = \sum_{j=1}^{N} \epsilon_j
\frac{{\it i} g_j^2}{(p-p')^2-\mu_j^2},
\label{kerg}
\end{eqnarray}
}
where index $j$ enumerates different "exchange" terms, distinguished
by the
mass parameter, $\mu_j$, and strength, $g_j$.
Factor $\epsilon_j=\pm 1$
defines attractive ($\epsilon_i=+1$) or repulsive ($\epsilon_j=-1$)
type of the correspondent term. Being interpreted as contribution
of the lowest order diagram,
the terms with
negative value of $\epsilon_j$ require either
 antihermitian term in the interaction hamiltonian
(lagrangian) or fields with additional quantum numbers.
However, we consider the entire sum in the eq. (\ref{kerg})
as a convenient parametrization of the unknown full kernel of the BS
equation.
The convenience of such a
form of the kernel is that kernel is explicitly covariant and
contains only "field theoretical" elements, the free particle
propagators.
Accepting this way of action, we assume that parameters of the
kernel should be fixed to describe the experimentally known spectrum
of the system.

The free propagator $D(p_1,p_2)^{-1}$ of the form (\ref{twoprop})
and the kernel in the ladder approximation, (\ref{ladk})
or (\ref{kerg}), usually provide the basis for an
investigation of the BS equation.
The parameters for a phenomenological
adjustment of the theoretical framework are the coupling constants,
$g_j$, and
exchange masses, $\mu_j$.
The parameter $m$ in the propagator is referred to as "physical mass"
of the constituent and it is supposed to include effectively  the
self-energy corrections. In the case of the observable particles,
e.g.
nucleons, $m$ is the observable mass, otherwise, e.g. quarks, it has
rather ambiguous model-dependent value.

In order to study the spectrum of the BS equation we, first, fix the
frame of
calculation
as the rest frame of the system, where $P=(M,{\bf 0})$ and
$M$ is the mass of the system, "meson". Then, we
 perform the well-known
Wick rotation~\cite{wick}, which may later cause difficulties in the
calculation
of observables of some reactions, but
does not affect the spectrum of
the equation and simplifies the numerical analysis.
Under the Wick rotation the BS equation keeps the form (\ref{sbse})
with
the ladder kernel
{\begin{eqnarray}
{\cal K}_{ladder}(\mu,p,p') =
\frac{g^2}{(p-p')^2_E+\mu^2},\quad (p-p')_E^2 = (p_0-p'_0)^2 +({\bf
p}-{\bf
p'})^2,
\label{ladkr}
\end{eqnarray}
}
and propagator
{\begin{eqnarray}
D(p,M) =
{\left [ (p_0^2+m^2 + {\bf p}^2- M^2/4)^2+M^2 p_0^2
\right ]}.
\label{twopropr}
\end{eqnarray}}
We use the previous notation for the "rotated"  functions
$\psi$, ${\cal K}$
and $D$, and for the "Euclidean" momenta, $p$ and $p'$.

Next, we perform a partial wave
decomposition of the equation~\cite{scale}:
{\begin{eqnarray}
\frac{1}{(p-p')_E^2+\mu^2}&=&\frac{2\pi}{\mid {\bf
p}\mid\cdot\mid{\bf p'}\mid}
\sum_{L=0}^{\infty} \sum_{\lambda=-L}^{L} Q_L(\beta)
Y_{L\lambda}(\theta_p,\varphi_p)
Y_{L\lambda}^*(\theta'_p,\varphi'_p),\label{part1}\\
\psi(p,M)
&=&\frac{1}{\mid {\bf p}\mid}
\sum_{L=0}^{\infty}
\sum_{\lambda=-L}^{L}
\psi_L(p_0,\mid {\bf p}\mid,M)
Y_{L\lambda}(\theta_p,\varphi_p),
\label{part}
\end{eqnarray}
}
where we have already taken into account
that $\psi_L$ are independent on the projection of angular momentum,
$\lambda$;
the dimensionless parameter $\beta$ defined by the expression
$$
\beta = \frac{\mu^2 +{\bf p^2 + p'^2} + (p_0 -p'_0)^2}
{2\mid {\bf p}\mid\cdot\mid{\bf p'}\mid},
$$
and $Q_l$ are the Legendre functions of the second kind. For $l=0,1$
they are:
\begin{eqnarray}
Q_0(y) = \frac{1}{2}{\sf ln}\left (
\frac{y+1}{y-1}
\right ) , \quad\quad
Q_1(y) = \frac{y}{2}{\sf ln}\left (
\frac{y+1}{y-1}
\right ) - 1 .\label{legql}
\end{eqnarray}

Substituting (\ref{part1}) and (\ref{part}) into eq. (\ref{sbse})
and performing the angular integration, we arrive at a set of
the independent equations
for the partial amplitudes $\psi_L$, corresponding to the
states with the angular momentum $L$:
{\begin{eqnarray}
&&\psi_L (p_0,|{\bf p}|,M) =   D(p,M)^{-1}\int
\frac{dp'_0 d|{\bf p'}|}{(2\pi)^3}
{\cal K}^L(p_0,|{\bf p}|,p'_0,|{\bf p'}|)
\psi_L(p'_0,|{\bf p'}|,M),
\label{sbsem}\\[2mm]
&&\quad\quad\quad\quad\quad\quad\quad\quad {\cal K}^L(p_0,|{\bf
p}|,p'_0,|{\bf
p'}|) = g^2 Q_L(\beta).
\label{park}
\end{eqnarray}}
The amplitude of pure quantum state with defined angular momentum
$L$ and its projection $\lambda$ is then given by
{\begin{eqnarray}
\psi(p,M,L,\lambda) = \psi_L( p'_0,|{\bf
p'}|,M)Y_{L\lambda}(\theta_p,\phi_p).
\label{pure}
\end{eqnarray}}

The two-dimensional integral equations (\ref{sbsem}) are an exact
projection of the initial equation (\ref{sbse}) with (\ref{ladk}) and
(\ref{twoprop}).  In the case of the generalized kernel (\ref{kerg}),
there will be a sum over different terms on the right hand side of
eq. (\ref{sbsem}):
{\begin{eqnarray}
{\cal K}_{G}^L(p_0,|{\bf p}|,p'_0,|{\bf p'}|)
 = \sum_{j=1}^N \epsilon_j g_j^2 Q_L(\beta_j).
\label{parks}
\end{eqnarray}
}

To obtain the spectrum of the BS equation we need to solve the
eigenvalue problem for the bound state
mass $M$ at fixed set of  parameters
of the kernel.
=It is to be noted that the
BS equation is not linear in the mass $M$.
On the other hand it is linear in the
coupling constants squared, $g^2$. Therefore we consider an
equation of the form:
{\begin{eqnarray}
\psi_L (p_0,|{\bf p}|,M) =   \lambda \cdot D(p,M)^{-1}\int
\frac{dp'_0 d|{\bf p'}|}{(2\pi)^3}
{\cal K}_G^L (p_0,|{\bf p}|,p'_0,|{\bf p'}|)\psi_L(p'_0,|{\bf
p'}|,M),
\label{sbsel1}
\end{eqnarray}
}
or symbolically
{\begin{eqnarray}
\psi =   \lambda \cdot  \hat {\cal K} \psi,
\label{sbsel}
\end{eqnarray}
}
where we skip indicies $L$ for shortness.

Solving the
linear eigenvalue problem for $\lambda$
at fixed parameters of the kernel
and various values of the bound state mass, we can map $\lambda(M)$.
Then by inverting the mapping as $M(\lambda)$, we  will find
solutions of the
eigenvalue problem for $M$ with the kernel $\lambda\cdot \hat{\cal
K}$,
where factor $\lambda$ is trivially absorbed by
redefining the coupling constants.

Traditional methods to find solutions to the linear eigenvalue
problem corresponding to the integral equation (\ref{sbsel})
are based on the ideas of the classical Fredholm theory.
The basic idea here is to substitute integration by a summation
and then deal with a sufficiently large system
of  linear algebraic equations. The problem is reduced to
finding the eigenvalues of the matrix corresponding to the
integral operator $\hat {\cal K}$.
However, in the case of the covariant BS equation
such a program of action meets extra difficulties in view of the
double integration, which leads necessarily to
very large matrices, say $\sim 10^4 \times 10^4$ or even larger
in the case of spinor-spinor equation (with additional factor of $64$
if the parity is conserved by interactions and factor 256 if it does
not).

We use an alternate method based on the iteration of the BS equation
in the form (\ref{sbsem}). This method in the essence is similar to
 the Malfliet-Tjon method~\cite{maltjon}
 employed to solve the integral Schr\"odinger
equation in the nuclear physics (see also
discussion in ref.~\cite{stadler}).

\subsection{The iteration method}

 The iteration
 of any trial function, $\Phi^{(0)}$, with eq. (\ref{sbsem}),
is understood as
obtaining other function $\Phi^{(n)}$ using the algorithm
(we use the symbolic notations as in eq. (\ref{sbsel})):
{\begin{eqnarray}
\Phi^{(i+1)} =   \lambda \cdot  \hat {\cal K} \Phi^{(i)},
\label{it1}\\
\Phi^{(n)} =   \left [ \lambda \cdot  \hat {\cal K} \right
]^n\Phi^{(0)},
\label{it2}
\end{eqnarray}
}

In order to organize the iteration (\ref{it1})-(\ref{it2})
on the computer,  we need
an "integrator", corresponding to
the operator $\hat {\cal K}$. This means that we have to perform the
computer calculation of the double integral on the r.h.s. of the
equation with a defined kernel and any trial function, $\Phi^{(0)}$,
(we assume
good enough behavior of the function, such as absence
of singularities and vanishing at large arguments, $|{\bf p}| \to
\infty$
and $p_0 \to \pm \infty$).
In our particular calculations this integrator is organized as a
two dimensional Gauss integration with suitable mapping of variables.
Next, we would like to know
what happens with equation after
sufficiently large number, $n$, of iterations.

Let us assume that  solutions, $\psi_\alpha$,
of the equation (\ref{sbsel}) corresponding to the eigenvalues
$\lambda_\alpha$
belong to the complete system of functions\cite{foot2}.
 Therefore, the trial function can be expanded
as
{\begin{eqnarray}
\Phi^{(0)} =   \sum_{\alpha=0}^\infty A_\alpha \psi_\alpha.
\label{it3}
\end{eqnarray}}
Thus, the result of iteration (\ref{it1})-(\ref{it2}) is:
{\begin{eqnarray}
\Phi^{(n)} =  \sum_{\alpha=0}^\infty A_\alpha
\left [\frac{ \lambda}{\lambda_\alpha}  \right ]^n\psi_{\alpha}.
\label{it4}
\end{eqnarray}}
{}From last equation it is obvious that at sufficiently large $n = N$
all terms with $\alpha > 0$ are small compared to the ground state
term, $\alpha = 0$:
{\begin{eqnarray}
\lim\limits_{{n\to N}}\Phi^{(n)} \equiv   \Phi^{(N)} =
 C \cdot \psi_{0} + {\cal O}\left (\left [
\frac{ \lambda_0}{\lambda_1}
\right ]^N \right ) .
\label{it5}
\end{eqnarray}}
Therefore $N$ to be chosen to make the last term on the r.h.s. of
(\ref{it5})
to be negligibly small. Then,
comparing $\Phi^{(N)}$ and $\Phi^{(N+1)}$,  we find
the ground state eigenvalue, $\lambda_0$:
{\begin{eqnarray}
  \lambda_0 = \frac{\Phi^{(N)}  }{  \Phi^{(N+1)}}.
\label{it6}
\end{eqnarray}
}

This recipe to find the ground state eigenfunction and eigenvalue
works
nicely numerically and, of course, does not depend on the choice of
the
initial trial function.

Formula (\ref{it3}) also provides us with the possibility to find
higher levels on  $\lambda$. Indeed, taking the combination
of iterating functions:
{\begin{eqnarray}
\Phi^{(i+1)}  - \left [
\frac{ \lambda}{\lambda_0}
\right ] \cdot \Phi^{(i)} = \sum_{\alpha=1}^\infty
\left ( 1 - \frac{\lambda_\alpha}{\lambda_0}\right )A_\alpha
\left [\frac{ \lambda}{\lambda_\alpha}  \right
]^{(i+1)}\psi_{\alpha},
\label{it7}
\end{eqnarray}}
i.e. new trial function which does not contain admixture of the
ground state! It is easy to see that iterations of this function
give the eigenfunction $\psi_1$ of the first excited state,
$\lambda_1$,
similarly to the procedure for ground state. The same procedure,
in principle, can be organized for any desired level on $\lambda$.
Thus, the problem is solved. The only limitation is, of course,
accuracy of the numeric calculations. Calculation of the high
$\lambda$
require a precise calculation of all levels below, which
leads to substantial computer time.
No special numerical problems were find in calculating the
three lowest
levels for eq.~(\ref{sbsel}).
The numerical results are presented in the next section.

\subsection{Numerical results for iteration method}

To study numerically the capability of the method to solve the
integral BS equation we consider eq.~(\ref{sbsem}) with the model
kernel of the form (\ref{parks}). The parameters of the model are
presented
in Table~I. We refer to the constituent fields as "quarks" and their
 mass, $m$, is chosen to be similar to the mass of the $c$-quark. The
parameters of the
kernel are chosen to provide the typical density of the levels of the
$c\bar c$-bound state, the charmonium, not too far
from the limit of the spectrum $M_{lim} = 2m$.

\begin{center}
\begin{tabular}{|c|c|c|}
\hline
  coupling constants & $\epsilon_j $ & mass             \\
  $g^2_j/(4\pi), \quad GeV^2$     &              &$\mu_j$, GeV
\\
 \hline
\hline
  37800.0 & +1 & 0.10    \\
\hline
  37800.0 & -1 & 0.11   \\
\hline
  45.0 & -1 & 0.95  \\
\hline
  45.0 & +1 & 1.425   \\
\hline
\hline
 \multicolumn{3}{|c|}{$m = 1.5 $ GeV, }\\
\hline
\end{tabular}

\vspace*{5mm}
Table. I. The parameters of model ("scalar charmonium").
\end{center}

Results of a calculation of the spectrum, $\lambda(M)$, are
presented in Fig.~1 for the three lowest levels in the channels with
$L=0$ (S-states) and $L=1$ (P-states).
The physical spectrum $M(\lambda)$ can be obtained
crossing  the plot by the line $\lambda = \lambda_{phys}$.
For instance,
as  is shown in Fig.~1 at $\lambda_{phys} = 0.13$ the masses of the
lowest states are (with accuracy $\sim 0.5\%$):
\begin{eqnarray}
M(1S) = 1.265 \quad {\rm GeV};
\quad M(2S) =1.939 \quad {\rm GeV};\quad M(3S) = 2.251 \quad {\rm
GeV};
\label{m1} \\
M(1P) =1.751 \quad {\rm GeV} ;
\quad M(2P) =2.146 \quad {\rm GeV};\quad M(3P) = 2.385 \quad {\rm
GeV}.
\label{m2}
\end{eqnarray}

For illustartion, we pesent the amplitudes $\psi_0$,
corresponding the spectrum (\ref{m1}) in Fig.~2.
These amplitudes are shown as a functions of the
spatial momentum $|{\bf p} |$
and at $p_0 = 0$. We see that the type of radial excitations
is similar to the one for the Schr\"odinger equation.
However, in general case of the BS equation we
deal with the hyperradial excitations of two-dimensional
surface, $\psi_L (p_0,|{\bf p} |)$.

\section {Confinement for the Bethe-Salpeter equation}

\subsection{General discussion and non-relativistic
confining potentials}

The idea of confinement has different realization within different
theoretical approaches. The simplest intuitive picture is given
by the non-relativistic bound state formalism based
on the Schr\"odinger equation with a QCD inspared
phenomenological potential.
A system of two particles interacting
in a  non-confining potential, vanishing
as $r \to \infty$, has a spectrum of bound states with an upper
limit, $M_{lim} = 2m_q$, and a continuum above this limit.
The confinement is conventionally associated with the
infinitely rising linear part of the
full $q\bar q$-potential,
$V_l = \alpha r, \alpha > 0$,
which provides with the mass spectrum extending infinitely
beyond $M_{lim}$.
It is clear that this mechanism can not be directly
adopted by the covariant field theoretical
approaches, such as the BS formalism.

More relevant approach is based on the simultaneous analysis of the
BS,  Schwin\-ger-Dyson (SD) equations and Ward-Takahashi at the
lowest order
and with the model
gluon propagator~\cite{munczek,williams}. In particular,
the role of the analytical structure (structure of singularities)
 of the
quark propagators is discussed here. It is found that, with
certain choices of the model gluon propagator, the quark
propagator is an entire function
(function with no singularities)
at physical momentum of the quark.
This important property of the
quark propagator is considered as an indication
of confinement~\cite{qpr,williams,tw}.

These two examples present two essentially different pictures of
what is referred to as confinement. In the first case, the
confinement is
the two-body effect, i.e. quarks can not escape from each other
because of the interaction between them. In the second example,
confinement is attributed to the property of a single
quark, which can not propagate as a free particle.

The general formalism of the field theory suggests that the
two-body and the single particle phenomena are in a generic
relationship. So do the
approaches based on the BS and SD equations.
Our approach to a modeling of the confinement is in some sense
inverse to that of refs.~\cite{qpr,williams}. We, first,
construct the covariant kernel of the BS equation
which, we expect, would provide the confinement  and only then
study the modifications of the quark propagators involved in the
BS equation.

We start from a few unsophisticated observations
prompted by the non-relativistic picture:
\begin{enumerate}
\item In the non-relativistic limit
the covariant theory can be reduced to the formalism with the
Schr\"odinger equation, where we can expect the picture
of the confinement as interaction with non-vanishing
potential at $r\to \infty$ to be valid. The main
distinguishing feature of the spectrum here is
the existence of the bound states above the
two quark mass limit, $M_{lim}$.
\item It is not necessary to have an infinitely
rising potential, if we intend to discuss only a
few lowest levels in the
spectrum. More manageable potential, $V \to V_{\infty} >0$ at
$r\to \infty$, could be sufficient, if $V_{\infty}$ is large enough.
In this sense, we also refer to the constant potential as
a confining one.
\item The  potential in the momentum space
$V({\bf k})$ can be obtained as a non-relativistic
limit of the kernel ${\cal K}$, similar to
\begin{eqnarray}
V({\bf k}) = -\frac{g^2}{{\bf k}^2+ \mu^2}
\label{yuk}
\end{eqnarray}
if the kernel is of the form (\ref{ladk}). Important point here is
that
the non-relativistic form of the potential (\ref{yuk})
is of a field-theoretic origin.
\end{enumerate}

Basing on the last observation we expect that if we define
a way to construct a confining potential from the non-relativistic
field theory, then we can apply similar methods to
obtain the relativistic confining kernel.
Very often the following recipe is used.
The non-relativistic Yukawa potential in the coordinate space $V(r)$,
the Fourier transform of eq. (\ref{yuk}), is
\begin{eqnarray}
V({\bf r}) = -\frac{g^2}{4\pi}\frac{e^{-\mu r}}{r}.
\label{yukr}
\end{eqnarray}
The linear potential can be derived as
\begin{eqnarray}
V_l = \lim\limits_{\mu \to 0}
\left [ -\frac{\partial^2}{\partial \mu^2}V({\bf r})
\right ]=
 \lim\limits_{\mu \to 0}\frac{g^2}{4\pi}{r}{e^{-\mu
r}}=\frac{g^2}{4\pi}{r}.
\label{dyukr}
\end{eqnarray}
The the relativistic generalization is made by a Fourier transform to
the
momentum space and replacing the non-relativistic Yukawa potential,
(\ref{yuk}), by it relativistic analog:
\begin{eqnarray}
{\cal K} \propto
 \lim\limits_{\mu \to 0}
\left [ -\frac{\partial^2}{\partial \mu^2}
\frac{g^2}{ {k}^2- \mu^2}
\right ].
\label{dyukp}
\end{eqnarray}
Taking the limit in (\ref{dyukp}), one should exercise
great deal of care, since this leads to the appearance
of generalized functions in the kernel~\cite{grossmilana,vary}.

This recipe give us a guideline, however it is not
completely satisfactory, since (i) the kernel (\ref{dyukp})
(or the potential
(\ref{dyukr})) is not of the field-theoretic form
and (ii) it is not clear does the
direct use of operation (\ref{dyukp})
lead to the rising or, at least, non-vanishing interaction in the
four-dimensional space.

Intending to stay with our parametrization of the
kernel as a superposition of the
ladder terms, similar to (\ref{kerg}), we have to find an
appropriate  presentation
of the operation (\ref{dyukr}). Let us start  with a superposition
of the non-relativistic potentials:
\begin{eqnarray}
V({\bf r}) = \sum_j   \frac{C_j}{r}{\sf exp}[-\mu_j r]
= \sum_j   \frac{C_j}{r}{\sf exp}[-\mu \alpha_j r],
\label{syukr}
\end{eqnarray}
where $\mu$ provides the mass scale and $\alpha_j$
 are  dimensionless parameters. Then expanding the exponents
we get
\begin{eqnarray}
V({\bf r})
&=&\sum_j   {C_j}\left [
\frac{1}{r} -\mu \alpha _j + \mu^2\frac{\alpha_j^2 r}{2} -\ldots
\right ].
\label{syukr1}
\end{eqnarray}

{}From eq. (\ref{syukr1}) we see that, taking
the limit $\mu \to 0$ and correspondly adjusting the parameters $C_j$
and
$\alpha_j$, desired  non-vanishing behavior of the potential can be
provided. For instance, for constant potential, $V_c$, we have
\begin{eqnarray}
C_j = \mu^{-1}\tilde C_j;
\quad \sum_j \tilde C_j = 0; \quad \sum_j \alpha_j\tilde C_j \equiv
A_c\ne 0,
\label{p0}
\end{eqnarray}
where we need only two terms  to satisfy the conditions, i.e.
$j_{max}=2$.
For linear potential,$V_c$,
\begin{eqnarray}
C_j = \mu^{-2}\tilde C_j;
\quad \sum_j \tilde C_j = 0; \quad \sum_j \alpha_j\tilde C_j = 0;
\quad \sum_j \alpha_j^2\tilde C_j \equiv A_l\ne 0,
\label{p1}
\end{eqnarray}
where $j_{max}=3$.
The limit $\mu \to 0$ corresponds to the physical
picture of the superposition of
very light mass exchanges with slightly different masses and
large coupling constants.
Please, note that power of the non-vanishing term in the limit
 $\mu \to 0$ is solely controlled by the power of the $\mu$
in the denominator of the coefficients $C_j$.

Using a superposition of the Yukawa potentials in the momentum space,
we find (see Appendix~A):
\begin{eqnarray}
V_c ({\bf k}) &=&  A_c \delta^{(3)}({\bf k}),
\label{p0p}\\[3mm]
V_l ({\bf k}) &=& \left(
\frac{1}{2}-\frac{{\sf ln}2}{\pi}
\right ) A_l\delta^{(3)}({\bf k})\frac{\partial }{\partial k}.
\label{p1p}
\end{eqnarray}
Note that, for the linear potential, the Schr\"odinger equation
becomes
an integro-differential equation in momentum space.

\subsection{Confining kernel for the Bethe-Salpeter equation}.

We look for a confining kernel, ${\cal K}_{con}$,
 of the BS equation in the form of the superposition
of ladder kernels in the momentum space and  after the Wick rotation:
\begin{eqnarray}
 {\cal K}_{con }(k_E) &=&
\sum_j \frac{C_j}{k_E^2 + \alpha_j^2 \mu^2},
\label{conk}
\end{eqnarray}
where $ k_E^2 = k_0^2+{\bf k}^2$.

In spite of the
obvious similarity to the non-relativistic case, the
expression for the relativistic kernel, (\ref{conk}),
has essentially different properties in view of the larger
dimension of the space. Indeed, the ladder kernel in the
coordinate space is~\cite{kercoor}:
\begin{eqnarray}
 {\cal K}_{ladder} (R_E,\mu) = g^2 \mu \frac{{\sf K}_1(\mu R_E)
}{R_E},
\label{kerco}
\end{eqnarray}
where $R_E = ({\bf r}^2+t^2)^{1/2}$. The asymptote at small $R_E$
can be obtained as:
\begin{eqnarray}
\!\!\!\!\! \!\!\!\!\!\!\!\!\!\! \!\!\!\!\! &&
 {\cal K}_{ladder} (R_E\to 0,\mu) \sim g^2 \left \{
\frac{1}{R_E^2} \right .
\nonumber \\
\!\!\!\!\! \!\!\!\!\!\!\!\!\!\! \!\!\!\!\!&& \quad
\left .+ \frac{\mu^2}{2} \sum_{m=0}^{\infty} \frac{1}{m!(m+1)!}
\left ( \frac{\mu R_E}{2} \right )^m
\left [
{\sf ln}
\left (\frac{\mu R_E}{2} \right ) -\frac{1}{2}
\left (\psi (m+1)+\psi(m+2)\right)
\right] \right \},
\label{kerco1}
\end{eqnarray}
where $\psi(m)$ is the Euler's psi function.
The expansion (\ref{kerco1})
is different from the behavior of the non-relativistic
potentials. However, a procedure similar to that of
the non-relativistic case can be applied to cancel the
lowest order
terms in the expansion (\ref{kerco1}) and generate
a non-vanishing
kernel in the limit $\mu \to 0$, since at large $R_E$
there is the exponential suppression similar to the
one in the Yukawa
potential:
\begin{eqnarray}
 {\cal K}_{ladder} (R_E\to \infty,\mu) \sim g^2
\mu^{1/2}
\frac{{\sf exp}[-\mu R_E]}{R_E^{3/2}} .
\label{kerco2}
\end{eqnarray}
We can see that the direct use of operation (\ref{dyukp})
does not lead to the desired non-vanishing at $R_E\to\infty$ behavior
of
the kernel.

We study here the kernel with the lowest power in $\mu^{-1}$, which
as we
expect controls the asymptotic behavior at large $R_E$.
Analysis in the momentum space,
similar to  that of the non-relativistic case,
gives the following conditions for the coefficients
 $C_j$ (see Appendix~B):
\begin{eqnarray}
C_j = \mu^{-2}\tilde C_j;
\quad \sum_j \tilde C_j = 0;
\quad \sum_j \alpha_j^2\tilde C_j = 0;
\quad \sum_j \alpha_j^2{\sf ln}\alpha_j\tilde C_j \equiv A \ne 0,
\label{prel}
\end{eqnarray}
which can be satisfied explicitly for $j_{max}=3$.
Note, that the second condition
cancels the most singular terms, $\sim R_E^{-2}$, in the coordinate
space, (\ref{kerco1}), similar to the non-relativistic potentials.
This choice of conditions leads to a kernel of the form:
\begin{eqnarray}
{\cal K}_{con}(k) = - {(2\pi)^4}U^4\delta^{(4)}(k_E),
\label{krel}
\end{eqnarray}
where for simplicity we introduce new effective coupling constant
$U$,
which has dimension of mass, and sign is chosen to provide us
with a confining-like kernel, similar to the positive constant
potential
in the non-relativistic case.
Fourier transform of the kernel, (\ref{krel}),
is a constant, $\sim U^4$, in four dimension. This means that it does
not behave like a non-relativistic constant potential, for which we
expect behavior like $\sim \delta(t_0) \cdot constant$.
Therefore, the kernel (\ref{krel}) does not exactly correspond to
the non-relativistic constant potential and the effective constant,
$U^4$,
is not related to constant in such a potential.
Note, since the constant $U^4$ does not have
the direct physical meaning, the choice of the factor ${(2\pi)^4}$ is
arbitrary and it is made for further simplification of
formulae.

The form of the kernel, (\ref{krel}), in accordance with
our main idea, is considered as a special limiting case of the
sum of the ladder kernels (sum of one-boson exchanges), which
provides
explicit covariance of the kernel and connection with the
usual field-theoretic
constructions. (Note that this is a
valid form in the Euclidian space, whereas
the transition to the Minkowsky space is not defined.)
By itself the $\delta$-form of the kernel is not something
very unusual in studying of the bound states of the quarks. For
instance,
such a form
is considered as "regularized" form of the highly
 singular, $\sim k^{-4}$,
behavior of the gluon propagator~\cite{qpr} and is a basis of the
models for studying of
the SD equation for the quark propagator~\cite{munczek,williams}. In
particular,
this form of the gluon propagator leads to the quark propagator
without singularities along physical momentum. In the lowest order,
such a gluon propagator gives  the kernel of the BS
equation~\cite{munczek}.

Let us study the effect of the kernel, (\ref{krel}), in the BS
equation,
(\ref{sbse}), under the Wick rotation. The form of ${\cal K}_{con}$
allows an integration in the equation explicitly:
{\begin{eqnarray}
\psi (p,P) =  - \frac{{U^4}}{ D(p_1,p_2) }
\psi(p,P).
\label{1con}
\end{eqnarray}
}
{}From eq.(\ref{1con}) we find that the kernel, ${\cal K}_{con}$,
does not
allow for the bound state solutions of the BS equation. In this sense
we
can expect that this kernel, in effect, is similar to the constant
non-relativistic potential,
which along does not allows for
the bound states.
 If this is the case, we expect that adding this
kernel to the regular attractive kernel, i.e. like the one presented
in Section~2.3, Table~1, we will get a shift of the spectrum by
a constant.

The BS equation with the combined kernel, ${\cal K}_G+{\cal
K}_{con}$,
can be transformed to the usual form,
but with a modified two-body propagator:
{\begin{eqnarray}
\psi (p,P) =  \frac{1}{D(p_1,p_2) + U^4}\int
\frac{d^4 p'}{(2\pi)^4}{\cal K}_G(p,p',P)\psi(p',P).
\label{2con}
\end{eqnarray}}
New propagator in eq. (\ref{2con}) has different analytical
properties, compared to the initial free propagator under
the Wick rotation,
(\ref{twopropr}), and main
difference is that the
new propagator does not have singularities
at $M > 2m$, which were a signal of the limit of the spectrum at
this point. This is an indication that the {\em physical
spectrum} exists beyond this point.
However, without numerical analysis we are not able to
discuss the properties of the solutions at $M > 2m$.

Before to go to the numerics, let us discuss the possible
effects of the self energy corrections in the presence
of the interaction of the form (\ref{krel}).
Dealing with the kernel (\ref{krel}) as a sum of the
lowest order (ladder) kernels, we calculate the
one-loop self-energy corrections to the single
quark propagator,
$d(p)^{-1}$.
Integration over the loop is performed at imaginary
$p_0$ component of the four-momentum of the quark and
result can be analytically continued to any values of
$p_0$. For physical momentum we get:
{\begin{eqnarray}
d(p) = p^2 -m^2 +\frac{U^4}{p^2-m^2}.
\label{singprop}
\end{eqnarray}}
The propagator (\ref{singprop}) does not have singularities
for physical values of the momentum, $p$. This property of
the interaction of the form (\ref{krel}) has already been
established within model investigations of the behavior of
the quark propagator~\cite{williams,qpr}.
Now we see that
this single quark effect is in a generic relation with
the two-body confining interaction in the framework of the
BS equation. For our calculations it is also important that
the singularities of the modified two-body propagator
in  (\ref{2con})
with $D(p_1,p_2)=d(p_{1})\cdot d(p_{2})$
defined by eq. (\ref{singprop})
still
allow to perform the Wick rotation.

\subsection{Numerical investigation
of the Bethe-Salpeter equation with the confining kernel}

First we compare the meson spectra obtained with eq.~(\ref{2con})
with
 $U^4 = 0$ and $U^4 \ne 0$. For convenience we take
the same kernel, ${\cal K}_G$, as that in the section~2.3.
The main effect, we expect here, is the shift of the spectrum
beyond the limiting point $M_{lim} = 2m$.
In order to compare the
spectra, we calculate $\lambda(M)$ for two
lowest states in the channels with $L=0,1$.
This is enough
to make conclusions about (i) the position of the bound
states, (ii) limiting point of the spectra and (iii)
separation between ground and excited states.
Results of these
calculations are presented in the Fig.~3, groups of curves A
and B. The constant $U$ for the case
B is chosen to be of the typical energy scale
of the equation, $U=m$.

We find, indeed, that the masen spectrum of the equation with $U\ne
0$
is extended beyond the point $M_{lim}$. However, it
displays unusual behavior beyond the point, $M_{lim}^B$
(this point is shown on Fig.~3 by the arrow)
\cite{foot3}.
Non-monotonical behavior of the curves $\lambda(M)$
indicates some difficulties in the interpretation
of the corresponding solutions. The obvious difficulty is
the existence of two solutions with the same coupling
constant $\lambda$
and different masses, $M$,
(see e.g. discussions in refs.~\cite{scale}).
Therefore, we have to find a way to
isolate the only one physical solutions.
The solution of this problem is quite simple.
Calculating the normalization of the solutions
of the equation along $M$ we find that the
ground state
solutions beyond $M_{lim}^B$ have a negative norm, i.e.
are the {\em abnormal} non-physical solutions.
\cite{foot4}.
This observation gives us the selection rule to eliminate the
extra solutions.

Another problem with the solutions corresponding $U^4 \ne 0$
is that spectrum is not only shifted to larger
masses, but also the separation between levels
is drastically
increased (see Table~II).
 This effect make it a problem for a
phenomenological
use of the kernel of the form (\ref{krel}). Indeed,
if we intend to consider states above $M_{lim}$ we have to
take $U^4$ large enough to provide us with a new limit of the
spectrum, however this can be in conflict with the desired
separation between levels. One may try to adjust the parameters
of the remaining part of the full kernel, ${\cal K}_G$,
so as to have a reasonable density of levels. However our analysis
showed that
in the presence of the kernel, ${\cal K}_{con}$,
separation between levels  depends only weakly on the
kernel, ${\cal K}_G$.

\vskip 6mm

\begin{center}
\begin{tabular}{|c|c|c|c|c|}
\hline
 $\Delta$ M  &  A       &   B        &   C        &   D\\
 &$\lambda^A_{phys}=0.13$&$\lambda^B_{phys}=0.25$
&$\lambda^C_{phys}=0.44$&$\lambda^D_{phys}=1.44$\\
 \hline
 \hline
 1P - 1S     & 0.49~GeV & 0.77 ~GeV   & 0.48~GeV   & 0.48~GeV  \\
\hline
 2S - 1P     & 0.19~GeV & 0.47 ~GeV   & 0.16~GeV   & 0.17~GeV \\
\hline
  2P - 2S    & 0.21~GeV & 0.42 ~GeV   & 0.22~GeV   & 0.26~GeV  \\
\hline
\end{tabular}

\vspace*{5mm}
Table. II. The levels splitting in the
 "scalar charmonium" for different kernels .
\end{center}

It should be remembered that in the BS equation the addition of
interaction
of any kind to the full kernel cannot be
related linearly to the shift in the mass of the system. The same
interaction also "shifts" the masses of the constituents
through the self-energy corrections to the single-particle
propagators.
A self-consitent approach has to be adopted to
include both the self-energy corrections and
changes in the two-body interactions.

We take into account the self-energy correction, (\ref{singprop}),
 to the
quark propagator. This leads to a corresponding  modification
of the two-body propagator $D(p,P)^{-1}$, (\ref{twoprop}).
The resulting spectrum is presented in Fig.~3, group of curves C.
It is clear that for the
interval of masses $\sim 2 \div 3.5$~GeV the density of
levels "returns to normal", of the same value as in the case A (see
Table~II).
The value of constant $\lambda_{phys}^C = 0.44$ is chosen as an
example
giving the spectrum density close to the the one of the case A.
At smaller $\lambda_{phys}^C $ the separation of levels is even
smaller.

The examples of calculations, B and C, show that the
 kernel containing the part ${\cal K}_{con}$ is
indeed similar in its effect
to the non-relativistic potential in coordinate space, $V(r)$,
with $V(r\to \infty) \to V_{\infty}$.
where the positive real constant $V_{\infty}$ defines
the shift of the  bound state spectrum
 compared to
the case with no ${\cal K}_{con}$, the case A.
 From Fig.~3  another similarity
 with the non-relativistic constant potential is obvious,
this kernel gives only a limited number of states in the spectrum,
which
can be adjusted by varying the $\lambda_{phys}$.
For instance, for the case C with $\lambda^C_{phys}=0.2 $
there is only one bound state, $1S$, whereas with
$\lambda^C_{phys}=0.45$ there are four states in the S and P channels
(there can be other undetected states in channels with higher
$L=2,\ldots$).

On the other hand, taking account of the self-energy corrections,
corresponding to this type of interaction,
leads to the disappearance of the poles in the single quark
propagator. Therefore quarks cannot propagate other than being
bound in a bound
state. Apparently, this fact is not related to the infinitely
rising interaction between them, but rather to the
modification of the single quark properties.

To be sure that the picture is valid in a wide interval of the
constant $U^4$,
also the spectrum is calculated
for the case $U^4 = 4m^4$ and self-energy
corrections taken into account.
The result is shown in the Fig.~3, group of curves D.
That the picture is found to be similar to
the one of case C, but the spectrum is shifted up to even larger
$M_{lim}^D\sim 4$~GeV. The density of levels in the interval
 of masses $\sim 2.5 \div 4$~GeV is the same as those of cases
A and C (see Table~II). The value of constant
$\lambda_{phys}^C = 0.44$ is chosen again
to give the spectrum density close to the the one of the case A.

Note that in our model the constant $U$ can not be taken arbitrary
large, since all the calculations are performed in the lowest order.
The natural criteria on the maximum value of $U$ is that the
corrections
of the lowest order, $\sim U^4$,
must not be too large. For instance, a shift of the mass spectrum
with $U\ne 0$
should not be too big compare to the typical masses for the case with
$U=0$.

\section {Conclusions}

We have presented a method to find
the ground state and excited states of the two-body
Bethe-Salpeter equation
for a channel with
any quantum numbers. This method allows us to
solve the bound state problem without reduction
of equation to quasipotential form or any other
approximations.

Based on a qualitative analogy with the construction
of a non-relativistic potential with non-vanishing asymptote
at large distances, $r\to\infty$, we have proposed a
recipe to obtain the confining kernel for the Bethe-Salpeter
equation, parametrized in the form of a special limiting
case of a superposition of  ladder kernels.
We find that in the simplest case such a kernel
is proportional to  $\delta(k_E)$ in the Euclidean momentum
space, which corresponds to a constant kernel in the coordinate
space.

We have studied the effect of this kernel on the spectrum of
the Bethe-Salpeter equation,
when the usual attractive interaction is added.
It is found that this kernel is similar in its effect
to the non-relativistic potential in coordinate space, $V(r)$,
with $V(r\to \infty) \to V_{\infty}$.
The positive real constant $V_{\infty}$ gives
the scale that defines the limit of the
 bound state spectrum compared to
the sum of the constituent masses, $M < 2m + V_{\infty}$.
At the same time the self-energy corrections
remove the singularities from the propagators of the constituents,
i.e.  constituents do not propagate.
Combination of these features of the solutions allows
an interpretation  of this type of interaction as
a confining interaction.

The illustrative analytical and
numerical calculation are presented for a model of  massive
scalar particles with scalar interaction and
do not pretend to a phenomenological
application.
However, the developed formalism can be straightforwarly
adopted for the Bethe-Salpeter equation for the bound state of
two spinor fields and, therefore, can be used for the realistic
studies of the properties of the quark-antiquark systems, mesons.


 \section{Acknowledgments}

 The authors thank
 A. Maximov and Yousuf Musakhanov for useful discussions.
 One of us (A.U.) is grateful to F. Gross, J. Milana and
 A. Stadler for the interesting discussions and comments.
 The research is supported in part
 by the Natural Sciences and Engineering Research Council of Canada.

\vspace*{.5cm}

{ \Large \bf Appendix A. The non-relativistic confining
potentials
in the momentum space.}
 \setcounter{equation}{0}
\def\theequation{A.\arabic{equation}}
\noindent

 In order to establish the form of the potentials defining by the
 eq. (\ref{syukr}), (\ref{p0}) and (\ref{p1}), let us consider
 auxiliary integral, $I_a$:
\begin{eqnarray}
 I_a &=& \int \frac{d^3 {\bf k}}{(2\pi)^3} V_a (k) f({\bf k})
\label{b1}\\
&=&\int \frac{d\Omega}{(2\pi)^3}
\int\limits_0^{\infty} dk k^2  V_a (k) f(k, \Omega),
\label{b11}
\end{eqnarray}
where $f({\bf k})$ is any function for which we
assume a "good" behavior ($f \to 0$ when $k\to \infty$,
no singularities, existence of derivatives, etc.),
$a = c, l$ depending on which of conditions the (\ref{p0}) or
(\ref{p1})
is imposed on the potential and $V_a(k)$
 defined as a fourier transform of
eq. (\ref{syukr}):
\begin{eqnarray}
 V_a (k) &=&
4\pi\sum_j \frac{4\pi C_j}{k^2 + \alpha_j^2 \mu^2}.
\label{b2}
\end{eqnarray}
The limit $\mu \to 0$ is assumed and we take it later.

Using the
common condition $\sum C_j =0$, we rewrite eq. (\ref{b11}) as
\begin{eqnarray}
 I_a
&=& -4\pi\mu^2\int \frac{d\Omega}{(2\pi)^3} \sum_j C_j\alpha_j^2
 \int\limits_0^{\infty}\frac{ dk}{k^2 + \alpha_j^2 \mu^2}
f(k, \Omega).
\label{b3}
\end{eqnarray}

Integrating by parts the last integral in (\ref{b3}),
we get:
\begin{eqnarray}
 I_a
&=& 4\pi\mu \int \frac{d\Omega}{(2\pi)^3}\sum_j C_j\alpha_j
 \int\limits_0^{\infty} dk \;{\sf arctg}\left [ \frac{k}{\alpha_j
\mu} \right ]
 f'(k, \Omega),
\label{b4}
\end{eqnarray}
where $f'= \partial f /\partial k$.

Let us now consider integration over $k$ only.
These integrals on the
r.h.s. of eq. (\ref{b4}) can be split in two parts:
\begin{eqnarray}
 I_a
\propto \mu \sum_j C_j\alpha_j \left \{
 \int\limits_0^{\alpha_j\mu}  +
\int\limits_{\alpha_j\mu}^{\infty} \right \} dk
\;{\sf arctg}\left [ \frac{k}{\alpha_j \mu} \right ]
 f'(k, \Omega).
\label{b5}
\end{eqnarray}
We estimate the first integral by the mean value theorem:
\begin{eqnarray}
 \mu \sum_j C_j\alpha_j
 \int\limits_0^{\alpha_j\mu}   dk \ldots
 = \mu^2 \left(
\frac{\pi}{4}-\frac{{\sf ln}2}{2}
\right )\sum_j C_j\alpha_j^2 f'(\xi\alpha_j\mu, \Omega),
\label{b6}
\end{eqnarray}
 where $0\le \xi \le 1$.

The second integral is estimated using the expansion
${\sf arctg}x = \pi/2 -1/x+1/(3x^3)-\ldots$, which is
valid at $x\ge 1$. It can be shown that
the first term in this expansion
gives the leading contribution to the
full integral in the limit $\mu \to 0$:
\begin{eqnarray}
 \mu \sum_j C_j\alpha_j
\int\limits_{\alpha_j\mu}^{\infty}  dk
\ldots = -\frac{\pi}{2}\mu \sum_j C_j\alpha_j
 f(\alpha_j\mu, \Omega).
\label{b7}
\end{eqnarray}

Finally, taking the limit $\mu \to 0$ and
accounting for the conditions on the coefficients
$C_j$, we find
\begin{eqnarray}
 I_c &=&  A_c f({\bf 0}),
\label{b8}\\
 I_l &=&  \left(
\frac{1}{2}-\frac{{\sf ln}2}{\pi}
\right ) A_l f'({\bf 0}).
\label{b9}
\end{eqnarray}
These equations give us the potentials in the form (\ref{p0p}) and
(\ref{p1p}).

{ \Large \bf Appendix B. The relativistic confining
kernel
in the momentum space.}
 \setcounter{equation}{0}
\def\theequation{B.\arabic{equation}}
\noindent

 In order to establish the form of the kernel defined by
 eq. (\ref{conk}) at the limit $\mu \to 0$ and lowest
(but non-zero) degree
 of $\mu^{-1}$ in the $C_j$,
 let us consider
 the auxiliary integral, $I$:
\begin{eqnarray}
 I &=& \int \frac{d^4 {\underline k}}{(2\pi)^4} {\cal K}_{con}(k)
 f({\underline k})
\label{c1}\\
&=&\int \frac{d\Omega^{(4)}}{(2\pi)^4}
\int\limits_0^{\infty} dk k^3  {\cal K}_{con} (k) f(k, \Omega^{(4)}),
\label{c11}
\end{eqnarray}
where $\underline k$ is four-momentum in Euclidian space,
$k = ({\bf k}^2 + k_0^2)^{1/2}$, $\Omega^{(4)} $ is the
hyperangle
defining the orientation of the vector $\underline k$
in the four dimensional space;
$f({k})$ is an arbitrary function for which we
assume "good" behavior ($f \to 0$ when $k\to \infty$,
no singularities, existence of derivatives, etc.).
The limit $\mu \to 0$ is assumed and we take it later.
We are omitting all nonessential factors, such as $2\pi$, etc.,
in the following calculations.

Adding and subtracting expression $C_j/k^2$  to each
term in (\ref{conk}),
 we rewrite eq. (\ref{c11}) as
\begin{eqnarray}
 I
\propto && -\mu^2\int {d\Omega^{(4)}}\sum_j C_j\alpha_j^2
 \int\limits_0^{\infty}\frac{ dk \, k}{k^2 + \alpha_j^2 \mu^2}
f(k, \Omega^{(4)})
\label{c3}\\
&&+ \int {d\Omega^{(4)}}\sum_j C_j
 \int\limits_0^{\infty}\frac{ dk}{k^2}
f(k, \Omega^{(4)}). \nonumber
\end{eqnarray}
Th last term is cancelled by the condition
\begin{eqnarray}
\sum_j C_j = 0.
\label{cc1}
\end{eqnarray}

Let us now consider integration over $k$ only.
This integral in
r.h.s. of eq. (\ref{c3}) can be evaluated as:
\begin{eqnarray}
 I
\propto -\mu^2 \sum_j C_j\alpha_j^2 f(\alpha_j\mu, \Omega^{(4)})
 \int\limits_0^{\Lambda}
 \frac{dk \; k}{k^2+\alpha_j^2 \mu^2},
\label{c5}
\end{eqnarray}
since the function remaining under integration has
sharp maximum at $k=\alpha_j\mu$. We also introduce the cut-off
parameter $\Lambda$ to regularize formally
the
logarithmically
divergent integrals. At a later stage of calculation
we take the limit $\Lambda \to \infty$, but for the moment
it is enough to assume $\Lambda \gg \mu$. Performing the integration
we get
\begin{eqnarray}
 I
\propto -\mu^2 \sum_j C_j\alpha_j^2 f(\alpha_j\mu, \Omega^{(4)})
 \left [ {\sf ln}\left ( \frac{\Lambda}{\mu} \right ) -{\sf
ln}\alpha_j
\right ].
\label{c6}
\end{eqnarray}
Applying the condition
\begin{eqnarray}
\sum_j C_j\alpha_j^2 = 0,\quad \sum_j C_j\alpha_j^2{\sf ln}\alpha_j
\ne 0,
\label{cc2}
\end{eqnarray}
and taking the limit $\mu \to 0$, we get
\begin{eqnarray}
 I
\propto \mu^2 f(0)\sum_j C_j\alpha_j^2
 {\sf ln}\alpha_j,
\label{c7}
\end{eqnarray}
which proves the form of the kernel as eq. (\ref{krel}).


\newpage
\centerline{ \large \bf Figure captions}

\vskip 5mm
Figure~1. Mapping of the bound state
spectrum of the Bethe-Salpeter equation.
The physical states correspond to the cutting of the trajectories
$\lambda(M)$ by the line $\lambda = \lambda_{phys}$. The
three lowest trajectories for $L=0$ (solid curves) and
$L=1$ (dashed curves)
are presented for the parameters of the kernel from Table~I.
An example of physical cut is shown for $\lambda_{phys}=0.13$
and corresponding masses are presented in (\ref{m1})-(\ref{m2}).

\vskip 5mm

Figure~2. The amplitudes corresponding to the three lowest $S-$states
($L=0$) for the Bethe-Salpeter equation with interaction
from the Table~I. These amplitudes corresponding to the states with
masses eq. (\ref{m1}) and $\lambda_{phys}=0.13$. Curves: ground state
($1S$) -
solid; first excited  state ($2S$) -
dashed; second state ($2S$) -
dotted.

\vskip 5mm

Figure~3. Mapping of the bound state spectrum of
the Bethe-Salpeter equation
with different kernels. The group of curves, A, is the two
lowest state from the Fig.~1 (see caption). Solid curves
present the S-states and the dashed curves presents the P-states.
The groups B, C and D
are trajectories for the same kernel as the group A
plus the confining part, ${\cal K}_{con}$:
B and C with $U^4 = m^4$; D with $U^4 = 4m^4$.
The calculations of groups C and D also take into account the
self-energy
corrections.
Arrows along the $M-$axis show the limits of the corresponding
physical spectra.


\begin{thebibliography}{99}
\bibitem{foot2} We disregard here the refined mathematical questions,
 such as existence and uniqueness of the solutions.
\bibitem{foot3}The minima of the  different trajectories, $\lambda
(M)$,
of the same group can be slightly shifted relatively to each other.
We denote the position of the minimum  of ground state trajectory
 as $M_{lim}^B$.
\bibitem{foot4} The same is valid for all trajectories of the group B
beyond their minima.
\end{thebibliography}
\end{document}